\documentclass[twocolumn,prd,showpacs,showkeys]{revtex4-1}
\usepackage{amsmath,amssymb,pgf,graphicx,color,url}
\newcommand{\be}{\begin{equation}}
\newcommand{\ee}{\end{equation}}

\begin{document}
\begin{flushright}  
INR-TH/2014-013
\end{flushright}
\vskip -0.9cm
%--------------------------------------
\title{Blind search for radio-quiet and radio-loud gamma-ray pulsars
  with Fermi-LAT data}
\author{G.\,I.\,Rubtsov$^{a,b}$}
\email{grisha@ms2.inr.ac.ru}
\author{E.\,V.\,Sokolova$^{a}$}
\email{sokol@ms2.inr.ac.ru}
\affiliation{
$^a$Institute  for Nuclear Research of the Russian Academy of Sciences,
Moscow 117312, Russia\\
$^b$Novosibirsk State University, Pirogov street 2,
630090 Novosibirsk, Russia}

\begin{abstract}
The Fermi Large Area Telescope (LAT) has observed more than a hundred
of gamma-ray pulsars, about one third of which are radio-quiet, i.e. not
detected at radio frequencies. The most of radio-loud pulsars are
detected by Fermi LAT by using the radio timing models, while the
radio-quiet ones are discovered in a blind search. The difference in the
techniques introduces an observational selection bias and, consequently,
the direct comparison of populations is complicated. In order to produce
an unbiased sample, we perform a blind search of gamma-ray pulsations using
Fermi-LAT data alone. No radio data or observations at optical or
X-ray frequencies are involved in the search process. We produce a
gamma-ray selected catalog of 25 non-recycled gamma-ray pulsars found in a
blind search, including 16 radio-quiet and 9 radio-loud pulsars. This
results in the direct measurement of the fraction of radio-quiet
pulsars $\varepsilon_{RQ} = 64\pm 10\%$, which is in agreement with
the existing estimates from the population modeling in the outer
magnetosphere model. The Polar cap models are disfavored due to a lower
expected fraction and the prediction of age dependence. The age, gamma-ray
energy flux, spin-down luminosity and sky location distributions of
the radio-loud and radio-quiet pulsars from the catalog do not
demonstrate any statistically significant difference. The results
indicate that the radio-quiet and radio-loud pulsars belong to one and
the same population. The catalog shows no evidence for the radio beam
evolution.
\end{abstract}
\maketitle

\section{Introduction}

A new stage of the history of radio-quiet gamma-ray pulsars discovery
began with the launch of Fermi satellite on June 11, 2008. Prior to
that, the Geminga pulsar~\cite{Geminga} stayed, for a long time, as the only
identified gamma-ray pulsar without a detectable radio counterpart. The
Fermi Large Area Telescopy (LAT) data enabled to discover 34 more
gamma-ray pulsars with both time-differencing
technique~\cite{Abdo12,Parkinson69} and a novel semi-coherent
method~\cite{Pletsh1,Pletsch2}. At present, 77 non-recycled pulsars are
identified by Fermi LAT, 42 of which are radio-loud,
see~\cite{Kerr:2012xh,Caraveo:2013lra} for review.

There are two general classes of gamma-ray pulsar models, namely the
Polar cap~(PC) and the outer magnetosphere~(OM) models. The classes
are characterized by a location of the origin of high-energy
emission. In the PC model~\cite{Sturrock:1971zc} the gamma rays are
produced by electrons accelerated in the polar cap region near the
surface of the neutron star. In the second class of models, electrons
are accelerated in outer regions of the pulsar
magnetosphere~\cite{Cheng:1986qt} (see also \cite{Li:2011zh} for
recent magnetosphere modeling results). A description of radio-quiet
pulsars is different in two model classes. In PC models, gamma-ray and
radio beams are produced in the same region and co-directed. In these
models, pulsars are observed as radio-quiet only if existing radio
survey sensitivity is not sufficient to detect radio
emission~\cite{SturnerDermer:1996,Gonthier:2002}. In the OM gamma-ray
and radio beams naturally possess different geometry, which leads to
a geometrical explanation of radio-quietness~\cite{Perera:2013wza}.

There is still an open question whether both radio-quiet and
radio-loud pulsars belong to the same population of astronomical
objects. It was noted that radio-quiet fraction is lower for young
pulsars~\cite{Ravi:2010sm}. This observation may be interpreted as the
time evolution of the width of the radio
beam~\cite{Ravi:2010sm}. Still the selection effects may be important
as the radio-quiet pulsars are identified with a completely different
procedure than radio-loud. In order to exclude this possible bias in
population studies, we build here the gamma-ray selected catalog of pulsars
applying the same blind search procedure to all Fermi-LAT point
sources, independently on the information known from radio or optical
observation.

The paper is organized as follows. In Section~\ref{sec:data}, the
Fermi LAT data analysis is explained. The photon list is build for
each of the Fermi-LAT point sources. In Section~\ref{sec:method}, we
overview the search method used for the
blind search of pulsations. Further we estimate the threshold for the
value of $H$-test in order to exclude a single false detection with a
probability of at least 90\%. The gamma-ray selected pulsar catalog
and the discussion of the results are presented in
Section~\ref{sec:results}.

\section{Data}
\label{sec:data}

We use the Fermi LAT Pass 7 (V6) publicly available weekly all-sky
data for the period from 2008 August 4 to 2013 March 6, corresponding
to the mission elapsed time (MET) from 239557418 to
384261063\,s~\cite{FermiLAT,Fermi_webdata}. We select ``SOURCE'' class
events with energies from 100\,MeV to 300\,GeV and apply the standard
quality cuts using {\it Fermi Science Tools v9r27p1}. We require that
zenith angle and satellite rocking angle do not exceed $100^\circ$ and
$52^\circ$ correspondingly.

We use all 1861 point sources from the Fermi LAT Second Source Catalog
(2FGL)~\cite{2FGL} as the candidates for the search. No preselection
of the sources based on the known type and properties is performed. We
use coordinates of the sources from 2FGL, although for many
sources the position is known better from other observations. This
is a price we pay for the blindness of the search to all data, except
gamma-ray radiation. We note, however, that the eficiency of the pulsar
search grows substantially if one includes the scan over the sky
location~\cite{Pletsh1}. The position variation requires more
computational resources and therefore the search in this paper is
limited to fixed positions.

For each source we build a source model which includes 2FGL sources in
a $8^\circ$ radius circle, galactic and isotropic diffuse emission
components (version P7V6). We fit Fermi LAT events in a circle of
$8^\circ$ with a model by {\it gtlike} tool using unbinned likelihood
analysis. The probability for each photon to be originated from the
source under consideration is obtained by {\it gtsrcprob} tool. This
probability is used in the following analysis as a weight of the
event. For each source we keep 12000 events with the highest
weights. The barycentric corrections to photon arrival times are
applied with {\it gtbary} tool.

\begin{figure}
\includegraphics[width=0.48\textwidth]{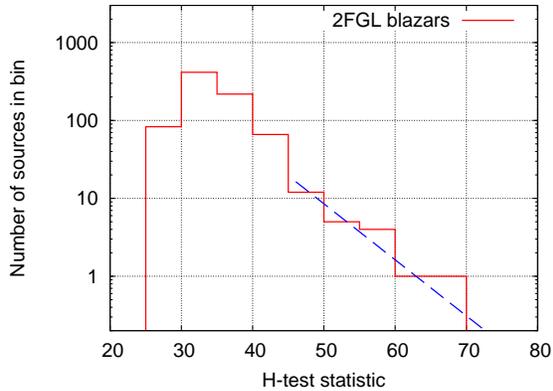}
\caption{The distribution of $H$-test statistic for 2FGL identified
  blazars. Dashed line represents an exponential fit of the
  tail.}
\label{histo}
\end{figure}

\section{Method}
\label{sec:method}

The search for pulsations is performed by summing up the 
spectral power over sub-intervals of time range. We scan over pulsar
frequency $f$ and spin-down rate $\dot{f}$ using the set of photon
barycentric arrival times $t_a$ and corresponding weights $w_a$. The
observation time range is split into $M=277$ intervals of length
$T=2^{19}$\,s, where the last interval is padded with zero flux
up to the length $T$. First, the arrival times, measured in MET
seconds, are corrected with

\begin{equation}
\label{time}
\tilde{t}=t + \frac{\gamma}{2}(t-t_0)^{2}\,,
\end{equation}
where $\gamma = \dot{f}/f$ and $t_0=286416002$\,s (MJD 55225) is a
reference epoch. We further bin each time interval into $N = 2^{25}$
bins and define $w_j^{(m)}$ as the sum of photon weights in the $j$-th bin
for the $m$-th time interval. Then, the spectrum $F_j^{(m)}$ in the $m$-th
interval is obtained with a discrete Fourier transform 
\begin{equation}
F_{k}^{(m)}=\sum_{j} w_j^{(m)} e^{2\pi i f_{k} \tilde t_{j}}\,,
\end{equation}
where $f_{k} = k/T$ and the Nyquist frequency is $N/2T = 32\,\mbox{Hz}$.
The Fourier transform is performed with the open-source Fast Fourier transform
library \textit{fftw}~\cite{fftw,fftw_web}.
Finally, the statistic $P_k$ is defined as a sum of squares of Fourier densities
over M time intervals
\begin{equation}
P_k = \sum\limits_{m=1}^{M} \left|F_k^{(m)} \right|^2\,,\\
\end{equation}

We scan over the parameter $\gamma$ from $0$ to $-1\times 10^{-12}$
with a step equal to $-2\times 10^{-15}$. The range corresponds to the
pulsar characteristic age greater than $16$\,kyr. The values of $f$ and $\dot{f}$
corresponding to the highest $P_k$ are then fine-adjusted by finding a
local maximum of the weighted $H$-test statistic~\cite{Htest}. The
latter is defined coherently on the whole time-interval as follows:

\begin{align}
\label{H}
H = \max_{1 \leq L
  \le20}\left[\sum_{l=1}^{L}\mid\alpha_{l}\mid^{2}-4(L-1)\right]\,,
\end{align}
where $\alpha_l$ is a Fourier amplitude of the $l$-th harmonic,
\begin{align*}
\alpha_{l} &= \frac{1}{\varkappa}\sum_{a}w_{a}\exp^{-2\pi i l f \tilde
  t_a}\,,\\
\varkappa^2 &= \frac{1}{2}\sum_{a}w_{a}^2\,.
\end{align*}

\begin{figure}
\includegraphics[width=0.48\textwidth]{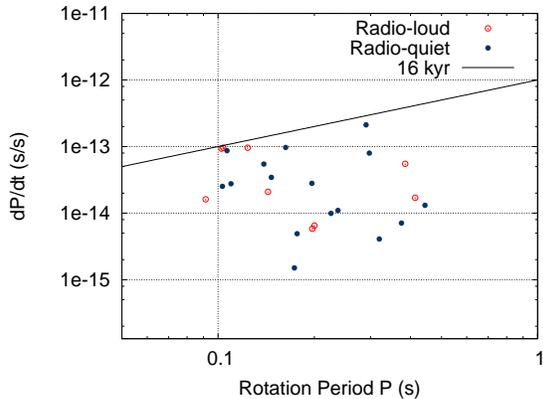}
\caption{ $P-\dot{P}$ plot for 25 pulsars found with a blind search in
  the present {\it Paper}. $P=\frac{1}{f}$ is a rotation period. The
  search is limited to characteristic ages $\tau_c > 16$\,kyr
  which correspond to the area below the solid line.}
\label{pp}
\end{figure}

We define the detection threshold $H_{th}$ in such a way that the
probability to have a single false candidate in the whole set does not
exceed $10\%$. As the distribution of $H$-test statistic for non-pulsating
objects is not known a priori, we construct an estimate for the
particular procedure of our scan. We produce a distribution of
$H$-test for 806 2FGL sources identified as blazars, see
Fig.~\ref{histo}. The tail of
the distribution is approximated with an exponential function and then
the value $H_{th}=83$ follows from the requirement that the integral of
extrapolated distribution above $H_{th}$ is equal to $0.1$. The above
background estimation technique is based on the complete scan
procedure and therefore naturally accounts for data-selection
and scans. Note
that the threshold determination required us to involve the sources
identification information, but this does not impact detection
uniformity as $H_{th}$ is a constant for all sources in the search.

\begin{table*}\small
\caption{\label{catalog} A catalog of gamma-ray pulsars found in a
  blind search. Frequency $f$ and spin-down rate $\dot f$ of gamma-ray
  pulsations correspond to the epoch MJD 55225. Age is estimated as
  $-f/2\dot f$. The last six columns contain the object information
  from the literature: pulsar name, galactic coordinates, Fermi LAT
  energy flux for $E>100$\,MeV\,\cite{2FGL}, type (Q - radio-quiet, L - radio-loud)
  and a reference to the first identification of gamma-ray
  pulsations. The improved pulsar positions
  from~\cite{Ray2011} are shown for the objects they are available.
}
%\begin{center}
\begin{tabular}{|c|c|c|c|c|c||c|c|c|c|c|c|}
\hline
\multicolumn{6}{|c||}{Blind search results} &
 \multicolumn{6}{c|}{Information from the literature}\\
\hline
no & 2FGL name & $H$-test & $f$, & $\dot{f}$, & age, & Pulsar & l, & b, & G, $10^{-11}$ & Type & Ref. \\
~ & ~ & ~ & Hz & $-10^{-13}$ Hz\,$s^{-1}$ & kyr & name & deg  & deg & erg\,cm$^{-2}$\,s$^{-1}$& ~ & ~\\ \hline 
1 & J0357.8+3205 & 2012 & 2.25172069 & 0.6645 & 537 & PSR J0357+32 & 162.76 & -16.01 & 6.5 & Q & ~\cite{Abdo12} \\ \hline
2 & J0633.7+0633 & 363 & 3.36242020 & 9.0032 & 59 & PSR J0633+0632 & 205.09 & -0.93 & 9.2 & Q & \cite{Abdo12} \\ \hline
3 & J0633.9+1746 & 3383 & 4.21755989 & 1.9515 & 343 & Geminga & 195.13 & 4.27 & 431.5 & Q & \cite{Geminga}   \\ \hline
4 & J0659.7+1417 & 118 & 2.59778597 & 3.7182 & 111 & Monogem pulsar & 201.11 & 8.26 & 2.5 & L & \cite{Ma4} \\ \hline
5 & J1028.5-5819 & 219 & 10.94042233 & 19.2759 & 90 & PSR J1028-5819 & 285.07 & -0.46 & 24.5 & L & \cite{Abdo:J1028-5819} \\ \hline
6 & J1044.5-5737 & 222 & 7.19264579 & 28.2455 & 40 & PSR J1044-5737 & 286.57 & 1.16 & 14.9 & Q & \cite{Parkinson69} \\ \hline
7 & J1048.2-5831 & 145 & 8.08368229 & 62.7803 & 20 & PSR B1046-58 & 287.42 & 0.58 & 20.5 & L & \cite{Thompson7} \\ \hline
8 & J1057.9-5226 & 7583 & 5.07321954 & 1.5032 & 535 & PSR B1055-52 & 285.98 & 6.65 & 29.3 & L & \cite{Thompson7} \\ \hline
9 & J1413.4-6204 & 149 & 9.11230488 & 22.9739 & 63 & PSR J1413-6205 & 312.37 & -0.74 & 16.4 & Q & \cite{Parkinson69} \\ \hline
10 & J1459.4-6054 & 139 & 9.69449976 & 23.7438 & 65 & PSR J1459-60 & 317.89 & -1.79 & 12.2 & Q & \cite{Abdo12} \\ \hline
11 & J1709.7-4429 & 2506 & 9.75607888 & 88.5384 & 17 & PSR B1706-44 & 343.11 & -2.68 & 135.1 & L& \cite{Thompson11}  \\ \hline
12 & J1732.5-3131 & 622 & 5.08792280 & 7.2595 & 111 & PSR J1732-31 & 356.31 & 1.01 & 21.2 & Q & \cite{Abdo12} \\ \hline
13 & J1741.9-2054 & 1269 & 2.41720730 & 0.9930 & 386 & PSR J1741-2054 & 6.43 & 4.91 & 12.2 & L & \cite{Abdo12} \\ \hline
14 & J1809.8-2332 & 936 & 6.81248059 & 15.9719 & 68 & PSR J1809-2332 & 7.37 & -2.01 & 49.3 & Q & \cite{Abdo12} \\ \hline
15 & J1836.2+5926 & 181 & 5.77154958 & 0.5004 & 1828 & PSR J1836+5925 & 88.88 & 25.00 & 60.3 & Q & \cite{Abdo12} \\ \hline
16 & J1846.4+0920 & 149 & 4.43357097 & 1.9517 & 360 & PSR J1846+0919 & 40.69 & 5.34 & 3.0 & Q & \cite{Parkinson69} \\ \hline
17 & J1907.9+0602 & 133 & 9.37779092 & 76.3568 & 19 & PSR J1907+06 & 40.18 & -0.89 & 28.2 & Q & \cite{Abdo12} \\ \hline
18 & J1957.9+5033 & 149 & 2.66804365 & 0.5040 & 839 & PSR J1957+5033 & 84.58 & 11.01 & 2.8 & Q & \cite{Parkinson69}  \\ \hline
19 & J1958.6+2845 & 408 & 3.44356138 & 25.1308 & 22 & PSR J1958+2846 & 65.88 & -0.35 & 9.5 & Q & \cite{Abdo12} \\ \hline
20 & J2021.0+3651 & 496 & 9.63902060 & 89.1185 & 17 & PSR J2021+3651 & 75.23 & 0.12 & 48.9 & L & \cite{Halpern}  \\ \hline
21 & J2028.3+3332 & 93 & 5.65907215 & 1.5721 & 571 & PSR J2028+3332 & 73.36 & -3.01 & 6.1 & Q & \cite{Pletsh1} \\ \hline
22 & J2030.0+3640 & 118 & 4.99678975 & 1.6230 & 488 & PSR J2030+3641 & 76.12 & -1.44 & 3.7 & L & \cite{Camilo22} \\ \hline
23 & J2032.2+4126 & 103 & 6.98089418 & 10.1823 & 109 & PSR J2032+4127 & 80.22 & 1.03 & 14.4 & L & \cite{Abdo12} \\ \hline
24 & J2055.8+2539 & 513 & 3.12928982 & 0.4005 & 1238 & PSR J2055+25 & 70.69 & -12.52 & 5.6 & Q & \cite{Parkinson69} \\ \hline
25 & J2238.4+5902 & 96 & 6.14486827 & 36.6124 & 27 & PSR J2238+5903 & 106.56 & 0.48 & 6.3 & Q & \cite{Abdo12} \\ \hline
%\end{center}
\end{tabular}
\end{table*}

\section{Results}
\label{sec:results}

\begin{figure}
\includegraphics[width=0.48\textwidth]{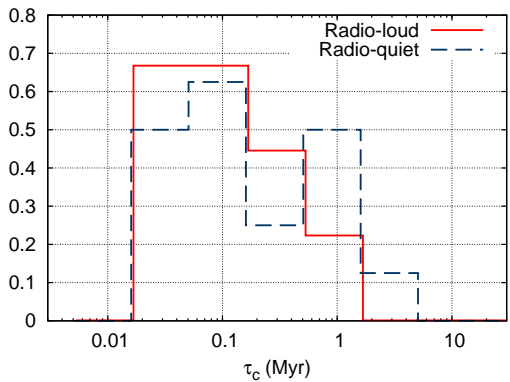}
\caption{Distributions of characteristic age $\tau_c=-\frac{f}{2\dot{f}}$
for radio-loud and radio-quiet pulsars. The two distrubutions are
compatible with KS probability $54\%$.}
\label{tau}
\end{figure}

We apply the procedure of Section~\ref{sec:method} to all 1861 point
sources of the 2FGL catalog. As a result, 25 objects are found with
the value of $H$-test above the threshold $H_{th}$, see
Table~\ref{catalog}. It appears that all of the pulsation detections
correspond to known gamma-ray pulsars, 16 of which are radio-quiet and
9 are radio-loud, see last six columns of Table~\ref{catalog}. This
allows us to estimate directly the fraction of radio-quiet pulsars
\begin{equation}
\varepsilon_{RQ} \equiv \frac{N_{RQ}}{N_{RQ}+N_{RL}} = 0.64\pm 0.10~\mbox{(68\%\,CL)}\,,
\end{equation}
where $N_{RQ}$ and $N_{RL}$ are numbers of radio-quiet and radio-loud
non-recycled gamma-ray pulsars. It should be mentioned that the
radio-quietness is determined according to the present-day sensitivity
of radio surveys. It is possible that faint radio emission will be
detected in the future from some of todays radio-quiet pulsars.

The fraction above confirms the domination of radio-quiet pulsars and
is perfectly consistent with the predictions of population synthesis
with OM model which gives a value of
0.65~\cite{Perera:2013wza}. The PC model with inverse Compton
gamma-ray production estimates the fraction as 
0.25~\cite{SturnerDermer:1996} which is excluded by the
observation. A curvature radiation version of the PC model leads to
the fraction value of 0.49-0.53~\cite{Gonthier:2002} which is slightly
disfavored. Moreover, the model~\cite{Gonthier:2002} is further
disfavored due to its prediction of the decrease of radio-quiet
fraction with age. Considering pulsars older than 100\,kyr the model
expectation of radio-quiet fraction 0.36 should be compared with
$0.62\pm0.13$ in our catalog.

The $P$-$\dot P$ plot for the pulsars from our catalog is shown in
Fig.~\ref{pp}. The distributions of characteristic age for radio-loud
and radio-quiet pulsars are compatible with the Kolmogorov-Smirnov (KS)
probability $54\%$, see Fig.~\ref{tau}. Therefore there is no
indication for the radio beam evolution. We note, hovewer, that the
pulsars younger that $\sim 16$\,kyrs are outside of the search range of the
present {\it Paper.}

\begin{table}
\caption{\label{tbl:KS} The KS-test probabilities for comparison of
  radio-quiet and radio-loud pulsar distributions over age,
  spin-down luminosity, energy flux above 100 MeV and galactic coordinates.}
\begin{tabular}{|c|c|}
\hline
Parameter & KS probability\\
\hline
age ($-f/2\dot f$) & 54\% \\
luminosity ($\sim f\dot f$) & 72\% \\
gamma energy flux & 43\%\\
l & 75\% \\
b & 69\% \\
\hline
\end{tabular}
\end{table}

The comparison of distributions over galactic coordinates, gamma-ray
energy flux and spin-down luminosity indicate that both radio-loud and
radio-quiet pulsars belong to the same population, see
Table~\ref{tbl:KS}. The agreement of the parameter distributions is an
additional argument in favor of the geometrical origin of radio-quietness
and therefore OM pulsar models are preferable. On the
contrary, the PC models unavoidably result in a strong age
dependence of the fraction of radio-quiet pulsars.

Given that Fermi LAT has by now observed 42 non-recycled radio-loud
pulsars, we expect according to our value of $\varepsilon_{RQ}$ that
there are about 75 radio-quiet pulsars among the Fermi-LAT
sources. The pulsed emission is discovered for only 35 of them leaving
$\sim 40$ sources as a challenge for the future pulsation
searches. In accord with our result, the machine-learning
classification of Fermi-LAT unidentified sources points to more than
50 gamma-pulsar candidates~\cite{LeeGuillemot:2012}.

\section*{Acknowledgments}

We are indebted to M. Pshirkov for numerous inspiring discussions. We
thank P. Tinyakov and S. Troitsky for helpful comments and
suggestions. GR is grateful for the hospitality of ULB Service de
Physique Theorique. The work was supported in part by the RFBR grants
12-02-31776, 12-02-91323 and 13-02-01293, by the Dynasty
foundation\,(GR), by the Russian Federation Government Grant
No. 11.G34.31.0047, by the grants of the President of the Russian
Federation NS-2835.2014.2, MK-1170.2013.2. The comparison of
population with the predictions of theoretical models is performed in
the framework of Russian Science Foundation grant 14-12-01340.  The
analysis is based on data and software provided by the Fermi Science
Support Center (FSSC). We used SIMBAD astronomical database, operated
at CDS, Strasbourg, France. The numerical part of the work is
performed at the cluster of the Theoretical Division of INR RAS.


\begin{thebibliography}{99}

\bibitem{Geminga} 
  J.~P.~ Halpern, S.~S.~Holt,  
  ``Discovery of soft X-ray pulsations from the gamma-ray source Geminga,''
  Nature\ {\bf357} (1992) 222-224.

\bibitem{Abdo12} 
  A.~A.~Abdo, et al.
  `` Detection of 16 Gamma-Ray Pulsars Through Blind Frequency Searches Using the Fermi LAT,''
  Science\ {\bf 325} (2009) 840. 

\bibitem{Parkinson69}
  P.~M.~Saz~Parkinson, et al.
  ``Eight gamma-ray Pulsars Discovered in Blind Frequency Searches of Fermi LAT Data,''
  Astrophys. J., {\bf 725} (2010) 571-584.

\bibitem{Pletsh1} 
  H.~J.~Pletsch, et al.
  ``Discovery of Nine Gamma-Ray Pulsars in Fermi-LAT Data Using a New Blind Search Method,''
  Astrophys. J. {\bf 744} (2012) 105.

\bibitem{Pletsch2}
H.~J.~Pletsch, et al.  
``PSR J1838–0537: Discovery of a young, energetic gamma-ray pulsar''
Astrophys. J., {\bf 755} (2012) L20.

\bibitem{Kerr:2012xh} 
  M.~Kerr [Fermi-LAT Collaboration],
  ``Pulsars in Gamma Rays: What Fermi Is Teaching Us,''
Proceedings of IAUS 291 "Neutron Stars and Pulsars: Challenges and
Opportunities after 80 years" (2012),
  arXiv:1211.3726.
  %%CITATION = ARXIV:1211.3726;%%
  %1 citations counted in INSPIRE as of 16 May 2014

\bibitem{Caraveo:2013lra} 
  P.~A.~Caraveo,
  ``Gamma-ray Pulsar Revolution,''
  Annual Review of Astronomy and Astrophysics vol. 52 (2014),
arXiv:1312.2913.
  %%CITATION = ARXIV:1312.2913;%%
  %1 citations counted in INSPIRE as of 14 May 2014

\bibitem{Sturrock:1971zc} 
  P.~A.~Sturrock,
  ``A Model of pulsars,''
  Astrophys.\ J.\  {\bf 164} (1971) 529.
  %%CITATION = ASJOA,164,529;%%
  %257 citations counted in INSPIRE as of 26 May 2014

\bibitem{Cheng:1986qt} 
  K.~S.~Cheng, C.~Ho and M.~A.~Ruderman,
  %``Energetic Radiation from Rapidly Spinning Pulsars. 1. Outer Magnetosphere Gaps. 2. Vela and Crab,''
  Astrophys.\ J.\  {\bf 300} (1986) 500.
  %%CITATION = ASJOA,300,500;%%
  %323 citations counted in INSPIRE as of 26 May 2014

\bibitem{Li:2011zh} 
  J.~Li, A.~Spitkovsky and A.~Tchekhovskoy,
  ``Resistive Solutions for Pulsar Magnetospheres,''
  Astrophys.\ J.\  {\bf 746} (2012) 60.
  %%  [arXiv:1107.0979 [astro-ph.HE]].
  %%CITATION = ARXIV:1107.0979;%%
  %38 citations counted in INSPIRE as of 26 May 2014

\bibitem{SturnerDermer:1996}
  S.J.~Sturner, C.D.~Dermer,
  ``Statistics of $\gamma$-ray pulsars.''
  A\&AS {\bf 120} (1996) 99.

\bibitem{Gonthier:2002}
P.L. Gonthier, M.S. Ouellette, J. Berrier, S. O'Brien, and
A. K. Harding,
 ``Galactic populations of radio and gamma-ray pulsars in the polar cap model,''
 Astrophys. J. {\bf 565} (2002) 482.

\bibitem{Perera:2013wza} 
  B.~B.~P.~Perera, M.~A.~McLaughlin, J.~M.~Cordes, M.~Kerr, T.~H.~Burnett and A.~K.~Harding,
  ``Modeling the non-recycled Fermi gamma-ray pulsar population,''
  Astrophys.\ J.\  {\bf 776} (2013) 61.
%  [arXiv:1309.1982 [astro-ph.SR]].
  %%CITATION = ARXIV:1309.1982;%%
  %1 citations counted in INSPIRE as of 14 May 2014

\bibitem{Ravi:2010sm} 
  V.~Ravi, R.~N.~Manchester and G.~Hobbs, 
  ``Wide radio beams from gamma-ray pulsars,''
Astrophys. J., 716 (2010) L85-L89.
% arXiv:1005.1966.
  %%CITATION = ARXIV:1005.1966;%%
  %1 citations counted in INSPIRE as of 16 May 2014

\bibitem{FermiLAT} 
  W.~B.~Atwood {\it et al.}  [LAT Collaboration],
  ``The Large Area Telescope on the Fermi Gamma-ray Space Telescope Mission,''
  Astrophys.\ J.\  {\bf 697} (2009) 1071.
%  [arXiv:0902.1089 [astro-ph.IM]].
  %%CITATION = ARXIV:0902.1089;%%
  %966 citations counted in INSPIRE as of 09 Apr 2014

\bibitem{Fermi_webdata}
  \url{http://fermi.gsfc.nasa.gov/ssc/data/access/}

\bibitem{2FGL} 
  Fermi-LAT Collaboration,
  ``Fermi Large Area Telescope Second Source Catalog,''
  Astrophys.\ J.\ Suppl.\  {\bf 199}, (2012) 31.
%  [arXiv:1108.1435 [astro-ph.HE]].
  %%CITATION = ARXIV:1108.1435;%%
  %420 citations counted in INSPIRE as of 02 Feb 2014

\bibitem{fftw}
M.~Frigo, S.G.~Johnson,
``The Design and Implementation of FFTW3,'' 
Proceedings of the IEEE {\bf 93} (2) (2005) 216;

\bibitem{fftw_web}
\url{http://www.fftw.org}

\bibitem{Htest}
O.C.~de Jager, B.C.~Raubenheimer, J.W.H.~Swanepoel,
``A poweful test for weak periodic signals with unknown light curve
shape in sparse data,''
A\&A {\bf 221} (1989) 180.

\bibitem{Ray2011}
P.S.~Ray et al., 
``Precise gamma-ray timing and radio observations of 17 gamma-ray pulsars,''
Astrophys.\ J.\ Suppl.\  {\bf 194}, (2011) 17.

\bibitem{Ma4}
  Y.~Ma, T.~Lu, K.~N.~Yu, C.~M.~Young 
  ``Possible discovery of three gamma-ray pulsars,''
  Astrophysics and Space Science {\bf 201} (1993) 113.

\bibitem{Abdo:J1028-5819} 
  A.~A.~Abdo {\it et al.}  [Fermi-LAT Collaboration],
  ``Discovery of Pulsed Gamma Rays from the Young Radio Pulsar PSR J1028-5819 with the Fermi Large Area Telescope,''
  Astrophys.\ J.\  {\bf 695} (2009) L72.
%  [arXiv:0903.1602 [astro-ph.HE]].
  %%CITATION = ARXIV:0903.1602;%%
  %26 citations counted in INSPIRE as of 21 May 2014

\bibitem{Thompson7}
D.~J.~Thompson 
``Gamma ray astrophysics: the EGRET results,''
Rep. Prog. Phys., {\bf 71} (2008) 116901.

\bibitem{Thompson11}
  D.~J.~Thompson, et al. 
``EGRET Observations of High-Energy Gamma Radiation from PSR B1706-44,''
  Astrophys. J., {\bf 465} (1996) 385.


\bibitem{Halpern}
  J.~P.~Halpern, et al.
 ``Discovery of High-Energy Gamma-Ray Pulsations from PSR J2021+3651 with AGILE''
  Astrophys. J., {\bf 688} (2008) L33

\bibitem{Camilo22}
  F.~Camilo, et al. 
  ``PSR J2030+3641: Radio Discovery and Gamma-Ray Study of a Middle-aged Pulsar in the Now Identified Fermi-LAT Source 1FGL J2030.0+3641''
  Astrophys. J., {\bf 746} (2012) 39.

\bibitem{LeeGuillemot:2012}
K.J.~Lee, L.~Guillemot, Y.L.~Yue, M.~Kramer, D.J.~Champion,
`` Application of the Gaussian mixture model in pulsar astronomy --
pulsar classification and candidates ranking for Fermi 2FGL catalog'',
MNRAS {\bf 424}, 4 (2012) 2832.

\end{thebibliography}
\end{document}